\documentclass[a4,12pt,german]{article}

\usepackage{pstricks,graphicx,eucal,mathrsfs}
\usepackage{amssymb,amsmath,amsxtra}

\newtheorem{criterion}{Criterion}
\newtheorem{theorem}{Theorem}

\newcommand{\be}{\begin{equation}}
\newcommand{\ee}{\end{equation}}
\newcommand{\eea}{\end{eqnarray}}
\newcommand{\bea}{\begin{eqnarray}}

\begin{document}


\title{{\bf \Large Quantum fields obtained from convoluted generalized white noise never have positive metric}}
\author{{\bf Sergio Albeverio$^1$} and {\bf Hanno Gottschalk$^2$}
\\[1ex] 
{\small $^1$Institut f\"ur angewandte Mathematik, Rheinische Friedrich-Wilhelms Universit\"at Bonn, Germany}\\{\tt \small albeverio@uni-bonn.de}
\\[1ex]
{\small $^2$Fachbereich f\"ur Mathematik und Informatik, Bergische Universit\"at Wuppertal, Germany}\\{\tt \small hanno.gottschalk@uni-wuppertal.de} \\}

\date{June 4$^{\mbox{th}}$ 2015}

\maketitle

{\abstract \noindent It is proven that the relativistic quantum fields obtained from analytic continuation of convoluted generalized (L\'evy type) noise fields have positive metric, if and only if the noise is Gaussian. This follows as an easy observation from a criterion by K. Baumann, based on the Dell'Antonio-Robinson-Greenberg theorem, for a relativistic quantum field in positive metric to be a free field. }

\vspace{.3cm}


\section{Introduction}
This note solves the problem, whether some of the Euclidean random fields constructed from convoluted generalized white noise \cite{AG,AGW1,AGW2,AGW3,AHK1,AHK2,AHK3,AIK,AW,BGL,GL,Osi,Sum} or white noise analysis \cite{GS} correspond to relativistic quantum fields acting on a (positive metric) Hilbert space.    

While it has been proven that such fields, analytically continued to Minkowski space-time, fulfill all Wightman axioms \cite{Jos,SW} except for, possibly, positivity, it has also been proven in \cite{AGW3} that they can always be represented on Krein spaces and thus fulfill the modified Wightman axioms by G. Morchio and F. Strocchi \cite{JS,MS,Stro}. Furthermore, in some cases they expose non trivial scattering behaviour \cite{AG,AGW4,AGW5,AGW6}.

A counter example for positivity is given in \cite{AGW2} for a special case. In this case, the Poisson part of the noise that drives the stochastic partial differential equation, from which the Euclidean random fields originate as a solution, is large compared to the Gaussian part. This however leaves the general case open, when the Poisson part of the driving noise is only a small or medium perturbation to the Gaussian one. See also \cite{Got1} for another partial non-positivity result in the case of presence of particles with masses $m_1$ and $m_2$ such that $m_1>2m_2$.

In this note we prove that relativistic quantum fields obtained by analytic continuation of the aforementioned random fields act on a positive metric Hilbert space, if and only if the Poisson part in the L\'evy noise is strictly zero. Despite the fact that the proof is an almost immediate consequence of a criterion by K. Baumann \cite{Bau} based on prior work by G. F. Dell'Antonio, O. W. Greenberg and D. W. Robinson \cite{DA,Gre2,Rob}, the problem has not been solved previously since the first publication of models from SPDEs driven by generalized white noise, see \cite{AHK1,AHK2,AHK3,AIK} for vector fields and \cite{AW} for scalar fields.  Since the criterion we use has only been formulated for scalar, charge free and fields, here we only consider the non positivity for the models described in \cite{AW,AGW1,AGW2,GS}.  With straight forward modifications of \cite{Bau,Rob}, the argument however also generalizes to the vector valued fields studied in \cite{AGW4,AGW5,AIK,AHK1,AHK2,AHK3,BGL,GL,Osi}. 

\section{Models from convoluted generalized white noise}  

In this section we shortly review some results of  \cite{AGW1,AGW2,AGW3,AW}. Let $d\geq 3$ be the space-time dimension\footnote{For the Dell'Antonio-Greenberg-Robinson theorem in $d=2$ there exist counter examples, see however \cite{Bau2}} and let $\eta$ be a generalized white noise with characteristic functional, in the sense of Bochner-Minlos \cite{Min}, given by 
\begin{equation}
\mathcal{C}_\eta(f)=\mathbb{E}[e^{i\eta(f)}]=\exp\left\{\int_{\mathbb{R}^d}\psi(f(x)) \, dx\right\},
\end{equation}
where $\psi(t)=ibt-\frac{\sigma^2}{2}t^2+\lambda \int_{\mathbb{R}\setminus\{0\}}(e^{ist}-1) dr(s)$ is a L\'evy characteristic with drift, diffusion and compound Poisson part and $f\in\mathcal{S}(\mathbb{R}^d,\mathbb{R})$ is a real-valued Schwartz function. We assume that $b\in \mathbb{R}$, $\sigma^2,\lambda\geq 0$ and $r$ a probability measure on $\mathbb{R}\setminus\{0\}$. Furthermore we assume that $r$ has moments of all orders, i.e. $r_n=\int_{\mathbb{R}\setminus\{0\}}s^n dr(s)<\infty$. 

Let us consider the stochastic pseudo differential equation for $m_0\geq 0$ and $0<\alpha<1$
\begin{equation}
(-\Delta+m_0^2)^\alpha\phi(x)=\eta(x)
\end{equation}
which is solved by the generalized random field $\phi(x)$ with characteristic functional
\begin{equation}
\label{eqa:ModEucl} 
\mathcal{C}_\phi(f)=\exp\left\{\int_{\mathbb{R}^d}\psi(G*f(x)) \, dx\right\},
\end{equation}
where $G(x)=\frac{1}{(2\pi)^d}\int_{\mathbb{R}^d}\frac{1}{(|k|^2+m_0^2)^\alpha}\, e^{ik\cdot x}\,dk$ is the Green function of the pseudo differential operator $(-\Delta+m_0^2)^{\alpha}$. $G*f(x)$ stands for the convolution of the tempered distribution $G$ with the test function $f$. We note that for $\alpha=\frac{1}{2}$ and $\lambda=0$ (i.e. pure Gaussian noise), we obtain the free field and for $0<\alpha<\frac{1}{2}$ and $\lambda=0$ a generalized free field is obtained \cite{AGW2}. For $\alpha>\frac{1}{2}$, the two point function is no longer (reflection) positive, hence this case is ruled out from the beginning. Suppose $\lambda>0$, then the truncated $n$-point Schwinger function, or $n$-th cumulant, of $\phi$ is
\begin{equation}
S_n^T(x_1,\ldots,x_n)=c_n\int_{\mathbb{R}^d}\prod_{j=1}^nG(x_j-y)\,dy
\end{equation}
with a constant $c_n=\lambda r_n>0$ if $\lambda>0$ and $n\geq 4$ is even. 
 
The Schwinger function can be analytically continued to the tube domains described in \cite{Jos,SW}. Taking the boundary values inside the tube to relativistic time, one obtains the corresponding Whightman functions $W_n^T(x_1,\ldots,x_n)$ with distributional Fourier transform are obtained in the case $0<\alpha<1$
\begin{equation}
\label{eqa:Wig}
\hat W_{n}^T(k_1,\ldots,k_n)=\tilde c_n \int_{R^n}\hat W_{n,m_1,\ldots,m_n}^T(k_1,\ldots,k_n)\prod_{l=1}^n\rho_{\alpha,m_0}(m_l^2)dm_1^2\cdots dm_n^2
\end{equation}
with $\rho_{\alpha,m_0}(m_l^2)=2\sin(\pi\alpha)\Theta(m^2-m_0^2)\frac{1}{(m^2-m_0^2)^\alpha}$ and
\begin{equation}
\label{eqa:WigM}
\hat W_{n,m_1,\ldots,m_n}^T(k_1,\ldots,k_n)=\left[\sum_{j=1}^n \prod_{l=1}^{j-1} \delta^+(k_l^2-m_l^2)\frac{1}{k_j^2-m_j^2}\prod_{l=1}^n \delta^-(k_l^2-m_l^2)\right]\delta\left(\sum_{l=1}^nk_l\right).
\end{equation} 
Here $k^2=(k^0)^2-\sum_{j=1}^{d-1}k_j^2$ is the invariant inner product on Minkowski space and $\delta^\pm(k^2-m^2)=\Theta(\pm k^0)\delta(k^2-m^2)$ is the invariant measure on the positive/negative $m$ mass shell. Here $\Theta(k^0)$ is the Heaviside function and $\frac{1}{k^2-m^2}$ is defined by its principal value with respect to $k^0$-integration. $\tilde c_n$ is some constant which, for $0<\alpha<1$, vanishes if and only if $c_n$ vanishes.

It has been proven \cite{AGW3,Got2,Stro} that for all these models, there exists a Hilbert space $\mathfrak{K}$ with positive definite inner product $(.,.)$, a self adjoint metric operator $\theta$ with $\theta^2=1$ and a vector $\Omega\in\mathfrak{K}$ with $\theta\Omega=\Omega$ such that
\begin{equation}
W_n^T(x_1,\ldots,x_n)=(\Omega,\Phi(x_1)\cdots\Phi(x_n)\Omega)^T.
\end{equation}
where $\Phi(x)$ is a operator valued distribution acting on some dense domain $\mathcal{D}\subseteq \mathfrak{K}$ which transforms covariantly under some $\theta$-unitary representation $U$ of the proper, orthochronous Poincar\'e group, i.e. $\theta U(g)\theta =U^{-1}(g)$ on $\mathcal{D}$ for $g\in \mathcal{P}_+^\uparrow(\mathbb{R}^d)$. The field operators are local in the usual sense $[\Phi(x),\Phi(y)]=0$ for spacelike separated $x$ and $y$ and the Wightman functions fulfill the cluster property. In the case of a mass-gap, this leads to uniqueness of the vacuum also in the indefinite metric case \cite{Got2}. Furthermore, the fields $\Phi(x)$ are $\theta$-symmetric, i.e. $\Phi^{[*]}(x)=\theta\Phi^*(x)\theta=\Phi(x)$. 

This boils the indefinite metric representation down  to a change of the involution from the usual involution $*$ on the Hilbert space $(\mathfrak{K},(.,.))$ to the involution $[*]$ on the inner product space $(\mathfrak{K},\langle.,.\rangle)$ where $\langle.,.\rangle=(.,\theta.)$ \cite{Got2,MS,Stro}. The positive metric case is thus equivalent to the case of a trivial metric operator, $\theta=1$.

\section{Baumann's criterion and non positivity for models from convoluted generalized white noise}

Based on the work in \cite{DA,Gre2,Rob}, K. Baumann proved the following criterion, when a local, chargeless and Bosonic quantum field with a unique vacuum is a free field: 

\begin{criterion}[\cite{Bau}] Let $\tilde S(\mathbb{R}^d,\mathbb{C})$ be the space of complex valued Schwartz functions $f$ with Fourier transform $\hat f(k)=\frac{1}{(2\pi)^{d/2}}\int_{\mathbb{R}^d} e^{ik\cdot x} f(x)\, dx$ such that the support of $\hat f$ is purely space-like, i.e. $\hat f(k)=0$ if $k^2\geq 0$. Let $W_n^T(x_1,\ldots,x_n)=(\Omega,\Phi(x_1)\cdots,\Phi(x_n)\Omega)$, $x_j\in\mathbb{R}^d$, denote the truncated Whightman functions of a scalar, chargeless quantum field $\Phi(x)$ that fulfills all Wightman axioms \cite{Jos,SW} including positivity. Suppose that for some $n\in\mathbb{N}$, $n\geq 2$, we have
\begin{equation}
\label{eqa:crit}
 W_{2n}^T(f,h_1,\ldots,h_{2n-2},g)=0 ~~\forall f,g \in S(\mathbb{R}^d,\mathbb{C}) \mbox{ and }h_j\in \tilde S(\mathbb{R}^d,\mathbb{C})
\end{equation}
then $\Phi(x)$ is a generalized free field in the sense of Greenberg \cite{Gre1}. Here $W_{2n}^T(f,h_1,\ldots,\linebreak h_{2n-2},g)$ stands for $W_{2n}^T(x_1,\ldots,x_n)$ smeared with the test functions $f,h_1,\ldots,h_{2n-2},g$.
\end{criterion}

Baumann's crterion can be traced back to the requirement that $[\Phi(h),\Phi(f)]\Omega=\linebreak \Phi(h)\Phi(f)\Omega=0$ for all $f\in S(\mathbb{R}^d,\mathbb{C})$ and $h\in \tilde S(\mathbb{R}^d,\mathbb{C})$ implies $[\Phi(h),\Phi(f)]\Omega=[W_2^T(h,f)-W_2^T(f,h)]\Omega$ for $f,h\in S(\mathbb{R^d},\mathbb{C})$ which is deduced from a representation of the Fourier transform of distributions vanishing for space-like arguments (like commutators do) with solutions of a $d+1$-dimensional wave equation \cite{Wig,Gre2,Rob}. Arguments of \cite{Wig} can then be used to enlarge the vanishing domain in momentum space up to the point that the scalar product $(\Psi,[ \hat \Phi(k_1),\hat \Phi(k_2)]\Omega)$ always vanishes, unless the projection of $\Psi$ to the zero momentum eigen space is non vanishing. Here $\hat\Phi(k)$ is the Fourier transform of $\Phi(x)$. But  the eigen space of zero momentum is spanned by the unique vacuum vector and one obtains $(\Psi,[\hat\Phi(k_1),\hat \Phi(k_2)]\Omega)=(\Psi,\Omega)(\Omega,[\hat \Phi(k_1),\hat \Phi(k_2)]\Omega)$ for all $\Psi$ in the Hilbert space. Using an inequality by K. Baumann \cite[Proof of Theorem 2, step b)]{Bau}, this implies canonical commutation relations  $[\Phi(h),\Phi(f)]=[W_2^T(h,f)-W_2^T(f,h)]$ with respect to the two point function $W_2^T$ and thus $\Phi(x)$ must be a free field, see e.g. \cite{Bau}.

We note that, although rightly remarked by N. Nakanishi and I. Ojima \cite{NO} the Greenberg-Robinson theorem also holds for quantum fields in indefinite metric, this is not true for Baumann's criterion, since, e.g. $W_4^T(f^*,h^*,h,f)=0$ does not imply $\Phi(h)\Phi(f)\Omega=0$ if the metric is indefinite and $h\in  \tilde S(\mathbb{R}^d,\mathbb{C})$ and $f\in S(\mathbb{R}^d,\mathbb{C})$. Here $f^*$ is the complex conjugate of $f$.

Baumann's criterion can be applied to prove non-positivity of the metric: suppose the criterion (\ref{eqa:crit}) holds for a given set of truncated Wightman functions that fulfill all Wightman axioms except, possibly, for the positivity axiom. Suppose, furthermore, that some  truncated Wightman function $W_l^ T(x_1,\ldots,x_l)$ with $l>2$ does not vanish identically and, for some $n\in\mathbb{N}$,  criterion (\ref{eqa:crit}) holds. Then the associated quantum field is not a generalized free field in the sense of \cite{Gre1}. Consequently, in order not to produce a contradiction with Baumann's criterion, the field can not fulfill the Wightman axiom of positivity. 

Let us thus check (\ref{eqa:crit}) for the truncated Wightman functions given in (\ref{eqa:Wig}) in Fourier space. As these are defined by integrals over the distributions $\hat W^T_{2n,m_1,\ldots,m_{2n}}$, it is sufficient to verify the condition for the latter. Using the Plancherel formula for tempered distributions, one obtains for $f,g \in S(\mathbb{R}^d,\mathbb{C})$ and $h_j\in \tilde S(\mathbb{R}^d,\mathbb{C}) $
\begin{equation}
\label{eqa:Check}
W_{2n}^T(f,h_1,\ldots,h_{2n-2},g)=\hat W_{2n}^T(\hat f,\hat h_1,\ldots,\hat h_{2n-2},\hat g).
\end{equation}
Using the explicit representation (\ref{eqa:WigM}), we see that each product of $\delta^\pm(k_l^2-m_l^2)$ with one of the functions $\hat h_l(k_l)$ vanishes as a consequence of the time-like support properties of $\delta^\pm(k_l^2-m_l^2)$ and the space-like support properties of $\hat h_l(k_l)$. However, in each $j$-term in the sum in (\ref{eqa:WigM}) there is at least one such pairing, since for  $n\geq2$ at least two functions $h_l(k_l)$ are present, but only one term $1/(k_j^2-m^2)$. Thus we see that the right hand side of (\ref{eqa:Check}) gives zero. We thus proved the following:

\begin{theorem}
None of the convoluted generalized white noise models obtained as the solution of (\ref{eqa:ModEucl}) for $\alpha>0$ has an analytic continuation to a relativistic quantum field in positive metric, unless $\lambda=0$ and $0<\alpha\leq \frac{1}{2}$. In particular, the Poisson part of the L\'evy characteristic has to be trivial. In this case, if $\alpha=\frac{1}{2}$, we obtain a free field and, if $0<\alpha<\frac{1}{2}$, a generalized free field.
\end{theorem}

Let us note that the above theorem analogously holds for the models by M. Grothaus and L. Streit \cite{GS} which share the basic structure of Wightman functions with the convoluted generalized white noise models and also for the models with non trivial scattering in \cite{AG} that scan be used to interpolate any given scattering amplitude. 

It is also straight forward to generalize the arguments of \cite{Bau,DA,Gre1,Rob} to Bosonic fields of general integer spin, as covariance under Lorentz transformations is only used to enlarge the space-like region in momentum space certain scalar products of the  quantum field vanish (and for the proof of the Reeh-Schlieder property, which however holds for arbitrary spin \cite{SW}). As in the criterion by Baumann (\ref{eqa:crit}), one assumes that truncated Wightman functions in Fourier space vanish for the entire spatial momentum region, this part of the argument is not even needed here. Therefore, our non positivity theorem equally holds for the vector fields in  \cite{AGW4,AGW5,AIK,AHK1,AHK2,AHK3,BGL,GL,Osi}.

\vspace{.5cm}

\noindent{\bf Acknowledgement:} We would like to thank the Bernoulli Centre Interfacultaire of EFPL for hosting us during this work in the framework of the Semester program "Geometric Mechanics, Variational and Stochastic Methods" 2015. In particular we thank the co-organisers A. B. Cruceiro and D. Holms.

\end{document}